\documentclass[prl,twocolumn,superscriptaddress,showpacs,preprintnumbers,amsmath2000,amssymb2000]{revtex4}
\usepackage{bm,graphicx,amstext2000,amsgen2000,amsbsy2000,amsopn2000}

\begin{document}

\title{Fluctuation spectrum of fluid membranes
coupled to an elastic meshwork:\\
jump of the effective surface tension at the mesh size}

\author{Jean-Baptiste Fournier}
\affiliation{Laboratoire de Physico-Chimie Th\'eorique, ESPCI, 10 rue
Vauquelin, F-75231 Paris cedex 05, France}
\affiliation{F\'ed\'eration de Recherche~CNRS~2438 ``Mati\`ere et Syst\`emes
Complexes",
F-75231 Paris cedex 05, France}
\author{David Lacoste}
\affiliation{Laboratoire de Physico-Chimie
Th\'eorique, ESPCI, 10 rue
Vauquelin, F-75231 Paris cedex 05, France}
\author{Elie Rapha\"el}
\affiliation{Laboratoire de Physique de la Mati\`ere
Condens\'ee, UMR
CNRS 7125, Coll\`ege de France, 11 Place Marcelin
Berthelot, F-75231
Paris cedex 05, France}
\affiliation{F\'ed\'eration de Recherche~CNRS~2438 ``Mati\`ere et Syst\`emes
Complexes",
F-75231 Paris cedex 05, France}

\begin{abstract}
We identify a class of composite membranes: fluid bilayers coupled to an
elastic meshwork, that are such that the meshwork's energy is a
function $F_\mathrm{el}[A_\xi]$ \textit{not} of the real microscopic
membrane area $A$, but of a \textit{smoothed} membrane's area $A_\xi$,
which corresponds to the area of the membrane coarse-grained at the mesh
size~$\xi$. We show that the meshwork modifies the membrane tension
$\sigma$ both below and above the scale $\xi$, inducing a tension-jump
$\Delta\sigma=dF_\mathrm{el}/dA_\xi$.  The predictions of our model
account for the fluctuation spectrum of red blood cells membranes
coupled to their cytoskeleton. Our results indicate that the
cytoskeleton might be under extensional stress, which would provide a
means to regulate available membrane area. We also predict an observable
tension jump for membranes decorated with polymer ``brushes''.
\end{abstract}


\date{\today}
\maketitle


Random surfaces~\cite{nelson_proceedings} play an important role in many
areas of physics, spanning from biophysics~\cite{lipowsky_book} and
chemical physics~\cite{safran_book} to high-energy
physics~\cite{wheater94}. In the last decades, there has been
considerable interest in the properties of fluid bilayer membranes
composed of amphiphilic molecules in water. It is
well-known~\cite{Helfrich73} that their fluctuations are dominated by
the membrane bending rigidity, $\kappa$, which leads to a spectrum
proportional to $(\kappa q^4+\sigma
q^2)^{-1}$~\cite{brochard75,engelhardt85,Meleard92}. Here, $\sigma$ is
an effective surface tension, i.e., an adjustable thermodynamic
parameter arising from the constraint of constant surface area, which is
usually several orders of magnitude smaller than the surface tensions of
ordinary liquid interfaces~\cite{Helfrich84,Evans90}. Additional
complexity arises if the membrane interacts with other systems (e.g., a
rigid substrate, another membrane, a network of polymers or
filaments)~\cite{evans97, joannic97, hackl98}, or contains inclusions
(passive or active)~\cite{goulian93,manneville99}.

In this paper, we identify a class of composite membranes, i.e., of
membranes coupled to an external system (like the cytoskeleton of red
blood cells~\cite{alberts_book}), for which the coupling energy can be
described, in a first approximation, by a function of the membrane area
\textit{coarse-grained} at a characteristic length scale $\xi$
(the mesh size in the cytoskeleton case). For such
systems, we show that the effective membrane tension should exhibit a
\textit{jump} of amplitude $\Delta\sigma$ at the scale $\xi$. In other
words, the fluctuation spectrum is proportional to $(\kappa
q^4+\sigma^<q^2]^{-1}$ for $q\lesssim\xi^{-1}$, and to $(\kappa
q^4+\sigma^> q^2)^{-1}$ for $q\gtrsim\xi^{-1}$. As we shall see,
$\Delta\sigma=\sigma^<-\sigma^>$ is directly related to the coupling
between the membrane and the external system. Understanding the scale
dependence of the effective surface tension, and in particular its value
at short length-scales, is important in a number of phenomena, e.g.,
membrane adhesion~\cite{servuss89,radler95}, cell
fusion~\cite{jahn02,yang02} and other microscopic biological mechanisms,
such as endocytosis~\cite{marsh99}. It should also help interpret
experiments on composite membranes~\cite{zilker87,strey95,safran02},
which attempt to determine the value of $\kappa$ by fitting the
fluctuation spectrum.

We shall consider two different systems for which our model predicts a
jump in surface tension: (i) membranes decorated with polymer
``brushes''~\cite{degennes87}, and (ii) red blood cell membranes
interacting with their attached cytoskeleton~\cite{alberts_book}.
Polymer decorated membranes are useful in providing a sterical
stabilization for liposome drug carriers~\cite{allen87,lasic95}; we
expect that better understanding their membrane tension should yield new
insight in their stability and mechanical properties. Here, we shall
mainly focus, however, on the composite membrane of the red blood cell
(or erythrocyte). The cytoskeleton of red blood cells is a roughly
triangular flexible protein meshwork, composed of spectrins attached to
the bilayer by association with integral membrane
proteins~\cite{alberts_book}. The resulting mechanical strength enables
red blood cells to undergo large extensional deformations as they are
forced through narrow capillaries. In the last two decades, the red
blood cell membrane has been thoroughly studied and a rich description
of its attached cytoskeleton has been
obtained~\cite{byers85,bennett89,mohandas94}.  Very recently, the
spectrum of the short wavelength membrane fluctuations of red blood
cells~\cite{zilker87}
has been successfully fitted by including the effect of confinement from
the cytoskeleton~\cite{safran02}; however, this approach leads to an
unexplained, rather abrupt change in the value of the membrane tension.
We shall show that our model explains quantitatively this change,
provided that the cytoskeleton is under (extensional) stress, which is
an important issue currently debated in cell
biology~\cite{boal98,boal_review}. Specific implications concerning red
blood cells as well as possible experiments on polymer decorated
membranes will be discussed at the end of the paper.

\begin{figure}
\includegraphics[width=.55\columnwidth]{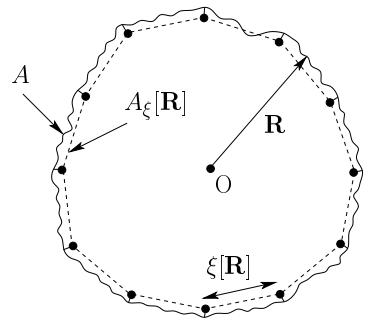}
\caption{Section of a fluctuating vesicle (thick line) to which
is attached a meshwork of springs (dashed lines), with common
length~$\xi$.  The actual membrane area is $A$ and the mesh area is
$A_\xi$, which corresponds to the membrane area coarse-grained at
$\xi$.}
\label{schema}
\end{figure}

Let us introduce our model by considering a membrane (embedded in
the Euclidian three-dimensional space) connected to an elastic
meshwork. The meshwork is composed of $N$ Hookean springs (spring
constant $\lambda$, free length $\ell_{0}$) and is attached to the
membrane through its nodes. The shape of the membrane is specified
by giving the position $\mathbf{R}(s^1,s^2)$ of a membrane element
as a function of two intrinsic coordinates. The total free energy
of the system may be written as
$F[\mathbf{R}]=F_{memb}[\mathbf{R}]+F_{el}[\mathbf{R}]$, where
$F_{memb}[\mathbf{R}]$ is the membrane free energy and
\begin{equation}\label{memb_elasticenergy}
F_\textrm{el}[\mathbf{R}]=N \frac{1}{2}\lambda
\left(\xi[\mathbf{R}]
-\ell_{0}\right)^2
\end{equation}
is the meshwork elastic energy. Since the meshwork cannot follow the
wiggles of the membranes of wavelength shorter than the mesh size
(see Fig.~\ref{schema}), the spring elongation $\xi[\mathbf{R}]$
actually depends on the membrane conformation. To simplify, we have
assumed that $\xi[\mathbf{R}]$ is identical for all the springs, which
is reasonable since the membrane is fluid.  The mesh size, i.e., the
spring length, can thus be expressed as a
\begin{equation}\label{consistency}
\xi^2[\mathbf{R}]=g\,\frac{A_\xi[\mathbf{R}]}{N},
\end{equation}
where $A_\xi[\mathbf{R}]$ is the area of the membrane
\textit{coarse-grained} at the mesh size, and $g$ is a constant
depending on the meshwork topology (e.g., $g=2\sqrt{3}$ for triangular
meshwork). Determining $\xi[\mathbf{R}]$ is the two-dimensional
analogous of measuring the length of a fractal coast (e.g., the coast of
Brittany~\cite{mandelbrot_book}) with a straight ruler, and determining
the ruler length that allows to go through the whole coast in $N$ steps.

Solving Eq.~(\ref{consistency}) for $\xi$ in the general case of an
arbitrary membrane shape is a rather difficult task. In order to
simplify the problem, let us consider the case of a quasi-spherical
membrane of fixed volume $V$ and fixed area $A$. If $V$ is expressed as
$V=\frac{4}{3}\pi R_0^3$, the area can then be written in the form
$A=\bar A+\delta A$ where $\bar A=4\pi R_0^2\gg\delta A$ is called the
projected area. We now consider a very small, almost planar portion
$\bar a=\epsilon\bar A$ of the projected area, that still contains a rather
large number $n=\epsilon N$ of springs. We represent the local membrane
shape by its fluctuation modes $u_\mathbf{q}$ above the local projected
area.  Locally, Eq.~(\ref{consistency}) can be expressed as
\begin{equation}
\frac{n}{g}\,\xi^2[u]=a_\xi[u]=
\bar a\,+\!\!\!\!\sum_{\,\,q\lesssim\xi^{-1}[u]}
\frac{1}{2}q^2|u_\mathbf{q}|^2+\mathcal{O}\left(u_q^4\right),
\end{equation}
where $a_\xi\approx\epsilon A_\xi$ is the local coarse-grained area.
Consistently with the second-order expansion, the upper limit
$\xi^{-1}[u]$ in the above sum can be replaced by its zeroth order
approximation:
\begin{equation}
\xi_0^{-1}=\sqrt{\frac{n}{g\bar a}}=\sqrt{\frac{N}{g\bar A}}.
\end{equation}
The elastic energy $f_\mathrm{el}=\frac{1}{2}n\lambda (\xi[u] -
\ell_{0})^{2}\approx\epsilon F_\mathrm{el}$ of the small portion under
consideration is then given by
\begin{equation}\label{felquad}
f_\mathrm{el}[u]=f_\mathrm{el}^{(0)}+ \Delta \sigma
\sum_{q\lesssim\xi_0^{-1}}\frac{1}{2}q^2|u_\mathbf{q}|^2+
\mathcal{O}\left(u_q^4\right),
\end{equation}
where
\begin{equation}\label{resultat_Delta_sigma}
\Delta \sigma = \frac{1}{2} g \lambda \left(1 -
\frac{\ell_0}{\xi_0}\right).
\end{equation}

\noindent
To proceed further, we introduce the standard form of the local membrane
elastic energy ($f_\mathrm{memb}\approx\epsilon F_\mathrm{memb}$), which
is given, to quadratic order, by~\cite{safran_book}
\begin{equation}\label{memb_energy}
f_\mathrm{memb}[u]=\sum_\mathbf{q}\frac{1}{2}\left( \gamma+\sigma
q^2+\kappa q^4
\right)|u_\mathbf{q}|^2.
\end{equation}
Here $\gamma$ is a confining potential, $\sigma$ an effective tension
yet to be determined (a Legendre transform multiplier constraining the total
membrane area), and $\kappa$ is the bending rigidity~\cite{Helfrich73}.
From the total energy $f_\mathrm{el}[u]+f_\mathrm{memb}[u]$, we deduce,
using equipartition, the fluctuation spectrum:
\begin{equation}\label{our_result}
\left\{
\begin{array}{lll}
\langle|u_\mathbf{q}|^2\rangle=
\displaystyle\frac{\epsilon\,k_\mathrm{B}T}
{\gamma+(\sigma + \Delta \sigma)q^2+
\kappa q^4}, & \text{for } & q\lesssim\xi_0^{-1},\vspace*{4pt}\\
\langle|u_\mathbf{q}|^2\rangle=
\displaystyle\frac{\epsilon\,k_\mathrm{B}T}
{\gamma+\sigma q^2+ \kappa q^4}, & \text{for } & q\gtrsim\xi_0^{-1}.
\end{array}
\right.
\end{equation}
The adjustable parameter $\sigma$ is determined from the requirement
$\sum_\mathbf{q}\frac{1}{2}q^2\langle|u_\mathbf{q}|^2\rangle=\epsilon\,\delta
A$. It follows that the decomposition of the low wavevector tension,
$\sigma^<$, and of the high wavevector tension, $\sigma^>$, in terms of
$\sigma$ and $\Delta\sigma$ ($\sigma^<=\sigma+\Delta\sigma$ and
$\sigma^>=\sigma$) is quite arbitrary: only
the values of $\sigma^<$ and $\sigma^>$ really matter.
Note that the position $\xi_0^{-1}$ of the jump only enters as a rough
characteristic scale in our model; a more refined calculation would be
required to determine the wavevector width of the jump. We have also
omitted the corrections to the bending rigidity originating from the
meshwork, which may also be scale-dependent.

The results obtained above are based on the fact that the meshwork
energy is actually a function of $A_\xi[\mathbf{R}]$ through Eqs.\
(\ref{memb_elasticenergy}) and~(\ref{consistency}). The same conclusions
concerning the effective tension therefore hold for \textit{any}
composite membrane for which the coupling may be approximated by a
function $F_\mathrm{el}(A_\xi)$ of the membrane area coarse-grained at
a characteristic scale $\xi$. The general expression giving the tension jump
occurring at $\xi_0\approx\xi$ is simply
\begin{equation}\label{general}
\Delta\sigma=
\left.\frac{\partial F_\mathrm{el}}{\partial
A_\xi}\right|_{A_\xi=\bar A}.
\end{equation}
Before making predictions concerning other systems, such as polymer
decorated membranes, let us discuss in more details the cytoskeleton
case.

\begin{figure}
\rotatebox{270}{\includegraphics[width=.6\columnwidth]{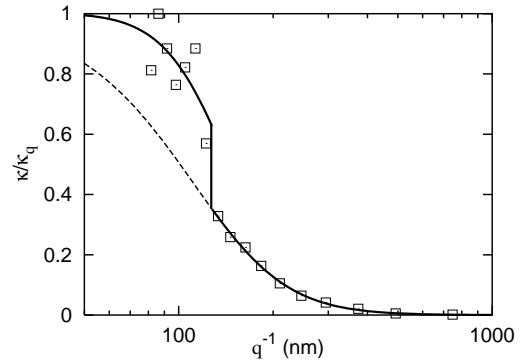}}
\caption{Fit (solid line) of the measured fluctuation spectrum of a
normal red blood cell (squares) from the data of Zilker et
al.~\protect\cite{zilker87}. The dashed line extrapolates the large
length-scale fit. The discontinuity and the corresponding tension jump
occurs at a length-scale $\simeq\!125~\mathrm{nm}$ comparable to the
cytoskeleton mesh.}
\label{fig:fit}
\end{figure}

Inspired by the recent work of Gov et al.~\cite{safran02}, we have
re-analyzed the data of Zilker et al.~\cite{zilker87} on the fluctuation
spectrum of normal red blood cells in the wavelength range
$300~\mathrm{nm}<\lambda<6000~\mathrm{nm}$. By fitting these data, the
authors of Ref.~\cite{safran02} have demonstrated the necessity to
introduce a confining potential, arising from the presence of the
cytoskeleton. They acknowledge, however, that the data at very short
scales are better fitted by a value of the membrane tension much smaller
than the one obtained at large scales.  Since the cytoskeleton is well
described by a meshwork of elastic
springs~\cite{alberts_book,boal_review}, we believe that the theory put
forward in this paper provides an explanation to this jump in surface
tension. Fluctuation spectra are usually fitted via an effective
wavevector-dependent bending rigidity $\kappa_q$ defined by
\begin{equation} \label{def kappa_q}
\langle|u_\mathbf{q}|^2\rangle=\frac{k_B T}{\kappa_q q^4}.
\end{equation}
According to Eq.~(\ref{our_result}), this quantity should roughly
exhibit a discontinuity at the mesh wavevector $q_0=\xi_0^{-1}$. This is
indeed the case, as evidenced by the solid line fit in
Fig.~\ref{fig:fit}. The position of the discontinuity being fairly
marked at $\xi_0^{-1}\simeq125~\mathrm{nm}^{-1}$, and the value
$\kappa\simeq2\times10^{-20}~\mathrm{N\,m}$ being already well
documented~\cite{safran02,strey95,brochard75}, we have fitted the data
to Eq.~(\ref{our_result}) for the parameters $\gamma$, $\sigma^<$ and
$\sigma^>$ only, obtaining
$\gamma=(5.2\pm0.6)\times10^6~\mathrm{J\,m}^{-4}$,
$\sigma^<=(1.5\pm0.2)\times10^{-6}~\mathrm{J\,m}^{-2}$ and
$\sigma^>=(-0.8\pm1.2)\times10^{-7}~\mathrm{J\,m}^{-2}$. According to
our model, $\Delta\sigma=\sigma^<-\sigma^>
\simeq1.6\times10^{-6}~\mathrm{J\,m}^{-2}$ should be given by
Eq.~(\ref{resultat_Delta_sigma}). Indeed, assuming $\xi_0>\ell_0$ (and
yet $\xi\approx\ell_0$) we deduce
$\lambda\approx10^{-6}~\mathrm{J\,m}^{-2}$, which agrees well with the
measured cytoskeleton elastic constants~\cite{lenormand01}. From the
positive sign of $\Delta\sigma$, our analysis predicts that the
cytoskeleton is under \textit{extensional} stress: this important issue
would require more experiments to test. Finding $\sigma^>$ essentially
zero, or even negative, is also quite remarkable. It suggests that one
important role of the cytoskeleton might be to regulate, via its elastic
stress, the membrane tension at length-scales shorter than the
cytoskeleton mesh, in order to produce either large fluctuations, or
even a short-scale buckling. This new type of area regulation (see
Ref.~\cite{morris01}) might prove useful to allow for large
deformations.

\begin{figure}
\includegraphics[width=.65\columnwidth]{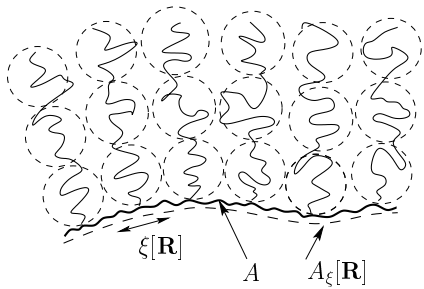}
\caption{Sketch of a polymer ``brush'' anchored on a fluctuating membrane.
The actual grafting density is the number of polymers divided by the
membrane area $A_\xi$ (dashed line) coarse-grained at the blob size $\xi$.}
\label{brosse}
\end{figure}

Another system to which our model can be applied is a membrane bearing
an anchored polymer brush. For a fixed number $p$ of polymers
in good solvent conditions, we have~\cite{degennes87}
\begin{equation}
\frac{F_\mathrm{el}}{p\,k_\mathrm{B}T}\simeq N
b^{5/3}(1-2\chi)^{1/3} \Sigma^{5/6},
\end{equation}
where $N$ denotes now the index of polymerization, $b$ the monomer
length, $\chi$ the Flory parameter, and $\Sigma$ the grafting density.
As illustrated in Fig.~\ref{brosse}, the relevant area defining the
grafting density is \textit{not} the total area $A$, but the area
$A_\xi$ coarse-grained at the blob scale. Hence,
\begin{equation}
\Sigma\simeq\frac{p}{A_\xi}\simeq\frac{1}{\xi^2}.
\end{equation}
Expressing as in Eq.~(\ref{felquad}) the brush energy in terms of the
local coarse-grained area, or equivalently, using Eq.~(\ref{general}),
we obtain
\begin{equation}
\label{jump_brush}
\Delta\sigma\simeq-N\,k_\mathrm{B}T\,b^{5/3}(1-2\chi)^{1/3}
\left( \frac{p}{\bar A} \right)^{11/6}.
\end{equation}
This actually corresponds to the brush osmotic pressure integrated over
the brush height:
$\Delta\sigma\approx-\Pi_\mathrm{osm}\,H\simeq-k_\mathrm{B}T(p/\bar
A)^{3/2} \times bN(b^2p/\bar A)^{1/3}$. Note that $\Delta\sigma<0$,
i.e., the brush reduces the large-scale tension with respect to the
short-scale one. It actually draws some of the membrane area stored at
length-scales shorter than the blob size in order to lower its grafting
density. Let us estimate
$\Delta\sigma$ in typical experiments~\cite{joannic97}: with $N=100$,
$b=1~\mathrm{nm}$, and a (projected) area per chain of
$100~\mathrm{nm}^2$, we obtain
$|\Delta\sigma|\approx10^{-4}~\mathrm{J\,m}^{-2}$, which compares with
ordinary membrane tensions. At length-scales larger than the brush
height $H$, the brush elasticity also yields a renormalization of
$\kappa$~\cite{hiergeist96}. Note that for distortions wavevectors in
the interval $H^{-1}<q<\xi^{-1}$, the brush will actually heal over
$\sim\!q^{-1}$, implying that the tension jump will probably be spread
over this interval~\cite{armand}. Our result agrees with a recent direct
calculation of Bickel et al.~\cite{bickel02}, showing that membranes
bearing polymers in the ``mushrooms" regime should possess an excess
tension $\simeq2k_\mathrm{B}T\,\Sigma$ at scales shorter than the
gyration radius $R_g$. Indeed, at the grafting density
$\Sigma\simeq1/R_g^2$ which crosses-over to the brush regime, this gives
$\Delta\sigma\simeq-\Pi_\mathrm{osm}H$, with $\Pi_\mathrm{osm}\simeq
k_\mathrm{B}T/R_g^3$ and $H\simeq R_g$. A possible way to test our
prediction Eq.~(\ref{jump_brush}) would be to measure, on a given
vesicle decorated with a polymer brush, the variation of the large-scale
membrane tension induced either by a modification of the solvent quality
or by a chemical alteration of the index of polymerization.

To conclude, the jump of the effective surface tension evidenced in this
paper arises from a competition between the free energy of the
membrane, which requires a certain amount of excess area to be stored in
the short wavelength modes, and the free energy of the coupled elastic
system, which is lowered if either less or more excess area (depending
on the system) is
transferred to the large-scale modes.  Very generally, a
wavevector-dependent membrane tension should therefore occur whenever an
external system coupled to the membrane produces a scale-dependent area
transfer.

We acknowledge fruitful discussions with A.~Ajdari, F.~Gallet, and
P.-G.~de~Gennes.

\end{document}